\documentclass[11pt]{article}

\usepackage[paper=a4paper,dvips,top=2.0cm,left=2.2cm,right=2.2cm,bottom=2.9cm,footskip=1.4cm]{geometry}
\usepackage{amsfonts,amsmath,amsthm,amssymb}
\usepackage{mathrsfs}
\numberwithin{equation}{section}  %%% Changes the equation numbering according to the section number
\usepackage[svgnames]{xcolor}
\usepackage{fix-cm}
\usepackage{sectsty}
\usepackage{fancyhdr}
\usepackage{lastpage}
\usepackage{graphicx}
\usepackage{url}
\usepackage{enumerate}
\usepackage{paralist}
\usepackage{multicol}
\usepackage{color}  %%% Needs to make colour of some content of your text
\usepackage{float}  %%% Fixes the position of the figure
\usepackage{epsfig}
\usepackage{enumitem}
\usepackage{hyperref}   %%% Needs to make hyper reference of any references or citations
\hypersetup{colorlinks=true,linkcolor=beamer@PRD, citecolor=beamer@PRD}
\usepackage{authblk}  %%% Helps to arrange author and institution details.
\usepackage{cite}  %%% Colour the citation brackets

\newcommand\myref[1]{\textcolor{beamer@PRD}{(}\ref{#1}\textcolor{beamer@PRD}{)}}

\definecolor{beamer@blue}{RGB}{0,0,255}
\definecolor{beamer@mediumblue}{RGB}{0,0,190}
\definecolor{beamer@midnightblue}{RGB}{25,25,112}
\definecolor{beamer@navy}{RGB}{0,0,128}
\definecolor{beamer@darkblue}{RGB}{0,0,139}
\definecolor{beamer@purple}{RGB}{128,0,128}
\definecolor{beamer@levander}{RGB}{100.,149.,237.}
\definecolor{beamer@PRD}{RGB}{46,48,146}
\definecolor{beamer@green}{RGB}{0,128,0}
\definecolor{beamer@darkgreen}{RGB}{0,100,0}
\definecolor{beamer@olive}{RGB}{128,128,0}
\definecolor{beamer@darkolivegreen}{RGB}{85,107,47}
\definecolor{beamer@gray}{RGB}{190,190,190}
\definecolor{beamer@ivry}{RGB}{220,220,220}%{238,232,205}
\definecolor{beamer@new}{RGB}{40,120,50}
\definecolor{shadecolor}{RGB}{220,220,220}
\definecolor{beamer@darkslategray}{RGB}{47,79,79}
\definecolor{beamer@chocolate}{RGB}{210,105,30}
\definecolor{beamer@brown}{RGB}{165,42,42}
\definecolor{beamer@orangered}{RGB}{255,69,0}
\definecolor{beamer@maroon}{RGB}{128,0,0}
\definecolor{beamer@white}{RGB}{234,242,243}
\definecolor{beamer@silver}{RGB}{0.5,0.45,0.37}

\begin{document}

%%%%%%%%%%%%%%%%%%%%%%%%%%%%%%%%%%%%%%%%%%%%%%%%%%%%%%%%%%%%%%%%%%%%%%%%%%%%%%%%%%%%%%%%%%%%%%%%%%%%%%%%%%%%%%%%%%%%%
%  Title
%%%%%%%%%%%%%%%%%%%%%%%%%%%%%%%%%%%%%%%%%%%%%%%%%%%%%%%%%%%%%%%%%%%%%%%%%%%%%%%%%%%%%%%%%%%%%%%%%%%%%%%%%%%%%%%%%%%%%%

\title{\textbf{Properties of soliton surfaces associated with integrable $\mathbb{C}P^{N-1}$ sigma models}}
\author{\textbf{Sanjib Dey$^{1,2,3}$ and A. M. Grundland$^{3,4}$} \\ \small{$^1$Institut des Hautes \'Etudes Scientifiques, Bures-sur-Yvette 91440, France \\ $^2$Institut Henri Poincar\'e, Paris 75005, France \\ $^3$Centre de Recherches Math{\'e}matiques, Universit{\'e} de Montr{\'e}al,
Montr{\'e}al H3C 3J7, Qu{\'e}bec, Canada\\ $^4$Department of Mathematics and Computer Science, Universit\'e du Qu\'ebec, \\ Trois-Rivi\`eres, CP 500 (QC) G9A 5H7, Canada \\ E-mail: sdey@ihes.fr, grundlan@crm.umontreal.ca}}
\date{}
\maketitle
    	
%%%%%%%%%%%%%%%%%%%%%%%%%%%%%%%%%%%%%%%%%%%%%%%%%%%%%%%%%%%%%%%%%%%%%%%%%%%%%%%%%%%%%%%%%%%%%%%%%%%%%%%%%%%%%%%%%%%%%%
%  Abstract
%%%%%%%%%%%%%%%%%%%%%%%%%%%%%%%%%%%%%%%%%%%%%%%%%%%%%%%%%%%%%%%%%%%%%%%%%%%%%%%%%%%%%%%%%%%%%%%%%%%%%%%%%%%%%%%%%%%%%%%%
\thispagestyle{fancy}
\begin{abstract}
We investigate certain properties of $\mathfrak{su}(N)$-valued two-dimensional soliton surfaces associated with the integrable $\mathbb{C}P^{N-1}$ sigma models constructed by the orthogonal rank-one Hermitian projectors, which are defined on the two-dimensional Riemann sphere with finite action functional. Several new properties of the projectors mapping onto one-dimensional subspaces as well as their relations with three mutually different immersion formulas, namely, the generalized Weierstrass, Sym-Tafel and Fokas-Gel'fand have been discussed in detail. Explicit connections among these three surfaces are also established by purely analytical descriptions and, it is demonstrated that the three immersion formulas actually correspond to the single surface parametrized by some specific conditions.
\end{abstract}	 
%%%%%%%%%%%%%%%%%%%%%%%%%%%%%%%%%%%%%%%%%%%%%%%%%%%%%%%%%%%%%%%%%%%%%%%%%%%%%%%%%%%%%%%%%%%%%%%%%%%%%%%%%%%%%%%%%%%%%%%
%  Introduction
%%%%%%%%%%%%%%%%%%%%%%%%%%%%%%%%%%%%%%%%%%%%%%%%%%%%%%%%%%%%%%%%%%%%%%%%%%%%%%%%%%%%%%%%%%%%%%%%%%%%%%%%%%%%%%%%%%%%%
\section{Introduction} \label{sec1}
\addtolength{\footskip}{-0.1cm} %% Adds extra space in the first page of the article
\addtolength{\voffset}{1.2cm} %% Adds extra space in the first page of the article
The construction of soliton surfaces associated with different integrable models has been an intense area of research over the last few decades. The most interesting and successful theory concerning this topic follows from the $\mathbb{C}P^{N-1}$ sigma models. Such models have a great importance in mathematical physics due to the reason that a significant number of physical systems can be reduced to relatively simple models defined either in Euclidean or in Minkowski space. In recent years, we have witnessed a rapid progress in the theory as well as its applications in several branches of modern science. One such promising example is in the context of quantum field theory and string theory, where the sigma models defined on spacetime and their supersymmetric extensions in Grassmannian manifold play essential roles in successful description of such theories \cite{Polchinski}. Other relevant applications of sigma models are found in numerous fields of physics; such as, in the Ising model of statistical physics \cite{Mccoy_Wu}, in the reduction of self-dual Yang-Mills field to the Ernst equation for cylindrical gravitational waves in gauge field theory \cite{Ablowitz,Nelson}, in the motion of boundaries between regions of different densities and viscosities in fluid dynamics \cite{Rozhdestvenski} and in different phenomena of condensed matter physics, e.g. the growth of crystals, deformation of membranes, dynamics of vortex sheets, surface waves, etc \cite{Nelson,David_Ginsparg}. In biochemistry and biology, soliton surfaces have been shown to play a crucial role in the study of biological membranes and vesicles, particularly in the study of long protein molecules \cite{Lipowsky,Davydov,Zhong} and the Canham-Helfrich membrane models \cite{Landolfi}. These macroscopic models can be derived from microscopic ones allowing one to explain the basic features and equilibrium shapes for biological membranes and the liquid interfaces \cite{Safran}. In chemistry, these theories are also applied in energy and momentum transport along a polymer molecule \cite{Davydov}.

The subject is very broad covering different areas of mathematical physics in conjunction with the surface theory. The growing number of new results, particularly from the experimental side motivates us to explore the models in a more compact way. Especially, by finding an explicit connection between the projectors and the generalized Weierstrass formula for the immersion of surfaces, we explore the possibility of expressing the $\mathbb{C}P^{N-1}$ models directly in terms of the Weierstrass surfaces. In addition, we discuss the links among several soliton surfaces existing in the literature in a complete analytical fashion. This will set up a direct connection of the theory of $\mathbb{C}P^{N-1}$ models with the surface theory studied in many different contexts giving more opportunity to the experimentalist to understand the subject in a more concrete way.

Our paper is organized as follows: In Sec. \ref{sec2}, we introduce the standard procedure for describing the $\mathbb{C}P^{N-1}$ models in terms of the rank-$1$ Hermitian projectors. Later, for the purpose of making our paper self-contained, we recollect some important properties of the projectors from the literature along with some new additions. In Sec. \ref{sec3}, we build new properties of the generalized Weierstrass surface by making use of the properties of the projectors. Relations among the Weierstrass surface and two other soliton surfaces, namely, the Sym-Tafel and the Fokas-Gel'fand are demonstrated in Sec. \ref{sec4}. Finally, our conclusions are stated in Sec. \ref{sec5}.  
%%%%%%%%%%%%%%%%%%%%%%%%%%%%%%%%%%%%%%%%%%%%%%%%%%%%%%%%%%%%%%%%%%%%%%%%%%%%%%%%%%%%%%%%%%%%%%%%%%%%%%%%%%%%%%%%%%%%%%
% Section 2
%%%%%%%%%%%%%%%%%%%%%%%%%%%%%%%%%%%%%%%%%%%%%%%%%%%%%%%%%%%%%%%%%%%%%%%%%%%%%%%%%%%%%%%%%%%%%%%%%%%%%%%%%%%%%%%%%%%%%
\section{$\mathbb{C}P^{N-1}$ sigma models in projector formalism in Euclidean space}\label{sec2}
\lhead{Properties of soliton surfaces associated with $\mathbb{C}P^{N-1}$ sigma models}
\chead{}
\rhead{}
\addtolength{\footskip}{0.1cm} %% Adds extra space in the first page of the article
\addtolength{\voffset}{-1.2cm} %% Adds extra space in the first page of the article
General properties of the $\mathbb{C}P^{N-1}$ sigma model and associated two-dimensional soliton surfaces in multidimensional Euclidean space have been studied extensively through many different techniques. The most interesting and fruitful approach follows from the description of the model in terms of the orthogonal rank-$1$ Hermitian projectors $P_k$,
\begin{equation}\label{ProjecProp}
P_k^2=P_k, \quad P_k^\dagger=P_k, \quad \text{tr}\left(P_k\right)=1, \quad P_lP_k=\delta_{lk}P_k, \quad \displaystyle\sum_{k=0}^{N-1} P_k=\mathbb{I}_N, \quad 0\leq (l,k)\leq N-1,
\end{equation}
which are defined on the Riemann sphere $S^2=\mathbb{C}\cup \{\infty\}$ with $\mathbb{I}_N$ being an identity matrix of dimension $N\times N$. The target space of the projectors $P_k$ are the complex lines in $\mathbb{C}^N$, which means that the projectors can be represented by the one dimensional vector functions $f_k(\xi,\bar{\xi})$ as given by
\begin{equation}\label{projector}
P_k=\frac{f_k\otimes f_k^\dagger}{f_k^\dagger\cdot f_k},
\end{equation}
where $f_k$ is a mapping $S^2 \ni (\xi,\bar{\xi})=x\pm iy\mapsto f_k\in\mathbb{C}^N-\{0\}$. Note that $P_k$ remains invariant when $f_k$ is multiplied by any scalar function. The relation \myref{projector}, in fact, provides an isomorphism between the equivalent classes of $\mathbb{C}P^{N-1}$ models and the set of rank-$1$ Hermitian projectors $P_k$. The projector formalism is sometimes more helpful in the sense that it automatically preserves the conformal and scaling symmetries and assures that the maps are free from removable singularities \cite{Goldstein_Grundland}, which might have occurred in the unnormalized vector fields $f_k$. Also, using this formalism, the equations of motion and related properties of the model become significantly compact, which we will discuss later in this section. Under the assumption that the action functional of the $\mathbb{C}P^{N-1}$ model is finite, the higher order rank-$1$ projectors $P_k$, as defined in \myref{projector}, can be obtained from the lowest order projector $P_0$, whose target space is an arbitrary holomorphic vector function $f_0(\xi)$, and vice versa, by using the following recurrence relations of $f_k$ \cite{Din_Zakrzewski,Goldstein_Grundland}
\begin{equation}
f_{k+1}=\left(\mathbb{I}_N-P_k\right)\cdot \partial f_k, \quad f_{k-1}=\left(\mathbb{I}_N-P_k\right)\cdot \bar{\partial} f_k,
\end{equation} 
where the holomorphic and anti-holomorphic derivatives are defined as
\begin{equation}
\partial\equiv\frac{\partial}{\partial\xi}\equiv\frac{1}{2}\left(\frac{\partial}{\partial x}-i\frac{\partial}{\partial y}\right), \qquad \bar{\partial}\equiv\frac{\partial}{\partial\bar{\xi}}\equiv\frac{1}{2}\left(\frac{\partial}{\partial x}+i\frac{\partial}{\partial y}\right).
\end{equation}
In terms of the projectors, the Lagrangian density and the Euler-Lagrange (EL) equations for the $\mathbb{C}P^{N-1}$ models \cite{Manton_Sutcliffe_Book} can be expressed as
\begin{equation}\label{EulerLagrnage}
\mathcal{L}_k=tr\left[\partial P_k\cdot\bar{\partial}P_k\right] \qquad \text{and}\qquad \left[\partial\bar{\partial}P_k,P_k\right]=\varnothing,
\end{equation} 
respectively, or equivalently in the form of a conservation law \cite{Zakharov_Mikhailov}
\begin{equation}\label{conservation}
\partial \left[\bar{\partial}P_k,P_k\right]+\bar{\partial}\left[\partial P_k,P_k\right]=\varnothing,
\end{equation}
with $\varnothing$ being a null matrix. The conservation law \myref{conservation} means that there exists a set of $1$-form
\begin{equation}\label{differential}
dX_k=i\left(\left[\bar{\partial}P_k,P_k\right]d\bar{\xi}-\left[\partial P_k,P_k\right]d\xi\right),
\end{equation}
which are closed differentials and can be utilized to construct the following $N\times N$ matrices in the form of contour integrals $\gamma_k$ in $\mathbb{C}$
\begin{equation}\label{EuSurface}
X_k\left(\xi,\bar{\xi}\right)=i\int_{\gamma_k}\left(\left[\bar{\partial}P_k,P_k\right]d\bar{\xi}-\left[\partial P_k,P_k\right]d\xi\right), \qquad X_k\left(\xi,\bar{\xi}\right)\in \mathfrak{su}(N),
\end{equation}
which may be identified as surfaces immersed in real $(N^2-1)$-dimensional Euclidean spaces \cite{Din_Zakrzewski,Grundland_Strasburger_Zakrzewski,Grundland_Yurdusen}. This mapping of an area of the Riemann surface $S^2$ into a set of $\mathfrak{su}(N)$ matrices, $S^2 \ni (\xi,\bar{\xi})\mapsto X_k(\xi,\bar{\xi})\in\mathfrak{su}(N)\simeq\mathbb{R}^{N^2-1}$, is known as the \textit{generalized Weierstrass formula for the immersion} of $2D$ surfaces in $\mathbb{R}^{N^2-1}$ \cite{Nomizu_Sasaki_Book,Konopelchenko,Konopelchenko_Taimanov}. By choosing a proper integration constant in such a way that the $X_k$'s become traceless, one obtains the expression of surfaces in an explicit form \cite{Grundland_Yurdusen}
\begin{alignat}{1}
X_k &=ic_k\mathbb{I}_N-i P_k-2i\displaystyle\sum_{j=0}^{k-1}P_j~\in\mathfrak{su}(N), \qquad c_k=\frac{1+2k}{N}, \label{surface} \\
&= i(c_k-1)P_k+i(c_k-2)\displaystyle\sum_{j=0}^{k-1}P_j+ic_k\displaystyle\sum_{l=k+1}^{N-1}P_l~\in\mathfrak{su}(N),
\end{alignat}
and, correspondingly, the inverse of \myref{surface} \cite{Goldstein_Grundland}
\begin{equation}\label{PkXk}
P_k=X_k^2-2i\left(c_k-1\right)X_k-c_k\left(c_k-2\right)\mathbb{I}_N.
\end{equation} 
One then computes the tangent vectors to the surfaces \myref{surface} as follows 
\begin{equation}\label{tangent}
\partial X_k=-i\partial P_k-2i\displaystyle\sum_{j=0}^{k-1}\partial P_j, \qquad \bar{\partial} X_k=-i\bar{\partial} P_k-2i\displaystyle\sum_{j=0}^{k-1}\bar{\partial} P_j,
\end{equation}
Also, from \myref{differential} the tangent vectors of the immersion are obtained as
\begin{equation}\label{TanXP}
\partial X_k=-i\left[\partial P_k,P_k\right], \qquad \bar{\partial} X_k=i\left[\bar{\partial} P_k,P_k\right],
\end{equation}
which when compared with \myref{tangent}, one obtains
\begin{equation}\label{comm}
\left[\partial P_k,P_k\right]=\partial P_k+2\displaystyle\sum_{j=0}^{k-1}\partial P_j, \qquad \left[\bar{\partial} P_k,P_k\right]=-\bar{\partial} P_k-2\displaystyle\sum_{j=0}^{k-1}\bar{\partial} P_j.
\end{equation}
\subsection{Characteristics of the projectors in $1D$-subspace}\label{subsec2.2}
Based on \cite{Goldstein_Grundland_1} together with some new findings, we list some important properties of the projectors $P_k$ in the following:
\begin{enumerate}[label=(\roman*)]
\item From the idempotent property $P_k^2=P_k$ \myref{ProjecProp}, it follows that the projectors $P_k$ are diagonalizable and their eigenvalues are $0$ and $1$. Since the projectors map to a one-dimensional subspace of $\mathbb{C}^N-\{\varnothing\}$, their rank is $1$. Therefore, only one of the eigenvalues of the projectors $P_k$ is $1$ and the rest is/are $0$. Hence $\text{tr}(P_k)=1$. Also, the idempotency of projectors $P_k^2=P_k$ implies that $(\mathbb{I}_N-2P_k)^{2k}=\mathbb{I}_N$ and $(\mathbb{I}_N-P_k)P_k=\varnothing$.
\item The differential consequences of $P_k^2=P_k$ are
\begin{alignat}{1}
& \left\{\partial P_k,P_k\right\}=\partial{P_k}, \quad \left\{\bar{\partial} P_k,P_k\right\}=\bar{\partial}{P_k}, \label{anticom}\\
& P_k\cdot\partial P_k\cdot P_k=P_k\cdot\bar{\partial} P_k\cdot P_k=\varnothing, \label{PDPP}
\end{alignat}
where $\{\cdot,\cdot\}$ represents the anticommutator between the matrices. Albeit, more general properties about the exchange among matrices $P_k,\partial P_k$ and $\bar{\partial}P_k$ in an arbitrary order, irrespective of the dimension of their target subspaces of $\mathbb{C}^N-\{\varnothing\}$ and projection angles, are given by
\begin{alignat}{1}
P_k\cdot\partial P_k\cdot\bar{\partial}P_k\cdot\cdot\cdot\cdot\cdot\partial P_k &= \partial P_k\cdot\bar{\partial} P_k\cdot\cdot\cdot\cdot\cdot\partial P_k\cdot P_k, \label{eqn216}\\
P_k\cdot\partial P_k\cdot\bar{\partial}P_k\cdot\cdot\cdot\cdot\cdot\partial P_k &= \partial P_k\cdot\bar{\partial} P_k\cdot\cdot\cdot\cdot\cdot\partial P_k\cdot \left(\mathbb{I}_N-P_k\right), \label{eqn217}
\end{alignat} 
for the total number of derivatives $\partial P_k$ and $\bar{\partial} P_k$ being even and odd, respectively.
\item Some simple consequences follow from \myref{ProjecProp} as
\begin{alignat}{1}
P_k\displaystyle\sum_{j=0}^{k}P_j &= P_k, \qquad P_j\displaystyle\sum_{l=0}^{N-2}P_l=P_j, \qquad j,l<k, \label{eqn218}\\
P_k\displaystyle\sum_{j=0}^{k-1}P_j &=\displaystyle\sum_{j=0}^{k-1}P_j\cdot P_k=\varnothing.
\end{alignat}
Using \myref{tangent} and by differentiating the first equation in \myref{eqn218} with respect to $\partial$ and $\bar{\partial}$, respectively, it is straightforward to compute
\begin{alignat}{2}
\partial P_k\displaystyle\sum_{j=0}^{k}P_j &= \partial P_k, \quad &&\displaystyle\sum_{j=0}^{k}P_j\cdot \partial P_k = P_k\cdot\partial P_k, \label{equation220} \\
\bar{\partial} P_k\displaystyle\sum_{j=0}^{k}P_j &= \bar{\partial} P_k\cdot P_k, \quad &&\displaystyle\sum_{j=0}^{k}P_j\cdot \bar{\partial} P_k = \bar{\partial} P_k, \label{equation220a}
\end{alignat}
which immediately imply to
\begin{alignat}{2}
\partial P_k\displaystyle\sum_{j=0}^{k-1}P_j &= P_k\cdot\partial P_k, \quad &&\displaystyle\sum_{j=0}^{k-1}P_j\cdot\partial P_k=\varnothing, \label{equation221} \\
\bar{\partial} P_k\displaystyle\sum_{j=0}^{k-1}P_j &=\varnothing, \quad &&\displaystyle\sum_{j=0}^{k-1}P_j\cdot\bar{\partial} P_k = \bar{\partial} P_k\cdot P_k. \label{equation221a}
\end{alignat}
The above Eqs. \myref{eqn218}--\myref{equation221a} have not been explored before, however, we will see that these simple relations are very helpful to reduce the complicated expressions to a compact form. 
\item From \myref{PDPP} it follows that
\begin{equation}
\text{tr}\left(P_k\cdot\partial P_k\right)=\text{tr}\left(P_k\cdot\bar{\partial} P_k\right)=0,
\end{equation}
and, consequently,
\begin{equation}
\text{tr}\left(P_k\cdot\partial^2 P_k\right)=-\text{tr}\left(\partial P_k\cdot\partial P_k\right),
\end{equation} 
and the analogous relations hold when we replace $\partial^2$ by the derivatives $\partial\bar{\partial},\bar{\partial}\partial$ and $\bar{\partial^2}$. Since, $P_k$ satisfies the EL equation \myref{EulerLagrnage}, by using \myref{ProjecProp} and \myref{PDPP} we have
\begin{alignat}{1}
& \text{tr}\left(\partial P_k\cdot\partial P_k\cdot P_k\right)=\text{tr}\left(\partial P_k\cdot\partial P_k\right) = 0 \label{eqn221}\\
& \text{tr}\left(\bar{\partial} P_k\cdot\bar{\partial} P_k\cdot P_k\right)=\text{tr}\left(\bar{\partial}P_k\cdot\bar{\partial} P_k\right) = 0 \label{eqn222}\\
& \text{tr}\left(P_k\cdot\partial P_k\cdot\partial\bar{\partial}P_k\right)=\text{tr}\left(P_k\cdot\bar{\partial} P_k\cdot\partial\bar{\partial}P_k\right) = 0, \label{eqn223}
\end{alignat}
for the proofs, see,\cite{Goldstein_Grundland_1}. Differentiating \myref{eqn221} and \myref{eqn222}, we obtain
\begin{equation}\label{equation224}
\text{tr}\left(\bar{\partial}\partial P_k\cdot\partial P_k\right)=0, \qquad \text{tr}\left(\partial\bar{\partial}P_k\cdot\bar{\partial} P_k\right)=0.
\end{equation}
It is assumed that $\partial\bar{\partial}P_k$ and $\bar{\partial}\partial P_k$ are continuous at any given point and, therefore, by using the Schwarz theorem for the mixed partial derivatives, one can say that $\partial\bar{\partial}P_k=\bar{\partial}\partial P_k$.
\item For any square matrix $A$ having the same dimension as the target space of the projectors $\mathbb{C}^N-\{0\}$, we get
\begin{equation}
P_k\cdot A\cdot P_k=\text{tr}\left(P_k\cdot A\right)P_k.
\end{equation}
\item The multiplication of any even number of $\partial P_k$ and $\bar{\partial} P_k$ matrices with at least one projector matrix $P_k$ yields
\begin{equation}\label{eqn225}
\partial P_k\cdot\bar{\partial} P_k\cdot\cdot\cdot\cdot\cdot\partial P_k\cdot P_k=\text{tr}\left(\partial P_k\cdot\bar{\partial} P_k\cdot\cdot\cdot\cdot\cdot\partial P_k\cdot P_k\right) P_k,
\end{equation} 
whereas the product of any odd number of identical $\partial P_k$ or $\bar{\partial} P_k$ matrices is given by
\begin{equation}\label{eqn226}
\partial P_k\cdot\partial P_k\cdot\cdot\cdot\cdot\cdot\partial P_k=\text{tr}\left(\partial P_k\cdot\partial P_k\cdot\cdot\cdot\cdot\cdot\partial P_k\cdot P_k\right) \partial P_k.
\end{equation}
\item From \myref{eqn221} and \myref{eqn222} we find some new and interesting properties, namely the left hand sides of \myref{eqn225} and \myref{eqn226} for the lowest possible number of identical derivatives are equal to zero, i.e.,
\begin{alignat}{1}
& \partial P_k\cdot\partial P_k\cdot P_k=\bar{\partial} P_k\cdot\bar{\partial} P_k\cdot P_k=\varnothing, \label{eqn227}\\
& \partial P_k\cdot\partial P_k\cdot\partial P_k=\bar{\partial} P_k\cdot\bar{\partial} P_k\cdot\bar{\partial} P_k=\varnothing, \label{eqn228}
\end{alignat}
and, thus, in general we have
\begin{alignat}{1}
P_k\cdot\partial P_k\cdot P_k\cdot\partial P_k\cdot\cdot\cdot\cdot\cdot\partial P_k\cdot P_k &= P_k\cdot\bar{\partial} P_k\cdot P_k\cdot\bar{\partial}P_k\cdot\cdot\cdot\cdot\cdot\bar{\partial}P_k\cdot\cdot P_k=\varnothing, \label{eqn229} \\
\partial P_k\cdot\partial P_k\cdot\cdot\cdot\cdot\cdot\partial P_k &= \bar{\partial} P_k\cdot\bar{\partial} P_k\cdot\cdot\cdot\cdot\cdot\bar{\partial} P_k=\varnothing, \label{eqn230}
\end{alignat}
for any number of $P_k$ (including $0$) in \myref{eqn229}, as well as, for any number of identical derivatives (greater than one in \myref{eqn229} except the case of $\partial P_k\cdot P_k\cdot\partial P_k(\neq \varnothing)$ and greater than two in \myref{eqn230}). The proof of \myref{eqn229} is obvious from \myref{eqn227} when we have only one $P_k$ on the rightmost position and the rest are derivatives. If $P_k$ is in the leftmost position and there are even number of derivatives on the right, then by using \myref{eqn216} $P_k$ can be brought to the rightmost position without any change. While for the case of odd number of derivatives by using \myref{eqn217} one obtains a factor $(\mathbb{I}-P_k)$ on the rightmost position. The result is zero in either case. In any other cases if we do not encounter the situation where the product contains factors like $P_k\cdot\partial P_k\cdot P_k=\varnothing$ or $\partial P_k\cdot\partial P_k\cdot P_k=\varnothing$, we can follow the same analysis as above to bring all of the $P_k$'s to the rightmost position, which always leave the product to be vanished. The only exception is $\partial P_k\cdot P_k\cdot\partial P_k(\neq \varnothing)$. It should be mentioned that with the introduction of \myref{eqn230} the property \myref{eqn226} becomes trivial.
\item The following traces vanish
\begin{equation}\label{eqn231}
\text{tr}\left(P_k\cdot\partial P_k\cdot P_k\cdot\bar{\partial}P_k\cdot\cdot\cdot\cdot\cdot\partial P_k\right) =0,
\end{equation}
where the total number of derivatives involving $\partial$ and $\bar{\partial}$ is odd in arbitrary order, as well as, the total number of projectors $P_k$ (including $0$) and their positions are also arbitrary. When the total number of derivatives $\partial$ and $\bar{\partial}$ is even, from \myref{eqn225} the relation turns out to be
\begin{equation}\label{eqn232}
\text{tr}\left(A\cdot\partial P_k\cdot\bar{\partial} P_k\cdot\cdot\cdot\cdot\cdot\partial P_k\cdot P_k\right)=\text{tr}\left(A\cdot P_k\right)~\text{tr}\left(\partial P_k\cdot\bar{\partial} P_k\cdot\cdot\cdot\cdot\cdot\partial P_k\cdot P_k\right),
\end{equation}
where $A$ is any square matrix of finite dimension $N$. If all the derivatives in \myref{eqn232} are either $\partial$ or $\bar{\partial}$, then by virtue of \myref{eqn229} and \myref{eqn230} the left hand side of \myref{eqn232} vanishes.
\end{enumerate}
%%%%%%%%%%%%%%%%%%%%%%%%%%%%%%%%%%%%%%%%%%%%%%%%%%%%%%%%%%%%%%%%%%%%%%%%%%%%%%%%%%%%%%%%%%%%%%%%%%%%%%%%%%%%%%%%%%%%%
%  Section 3
%%%%%%%%%%%%%%%%%%%%%%%%%%%%%%%%%%%%%%%%%%%%%%%%%%%%%%%%%%%%%%%%%%%%%%%%%%%%%%%%%%%%%%%%%%%%%%%%%%%%%%%%%%%%%%%%%%%%%%%
\begin{section}{Properties of the generalized Weierstrass formula for the immersion of surface}\label{sec3}
To explore the properties of the $\mathbb{C}P^{N-1}$ model in terms of the surface associated to it, in this section we prove several properties of the surfaces $X_k$, which are used in various branches of mathematical physics to a greater extent than the projectors.
\paragraph{Property 1:}
$X_k$'s belong to the set of $\mathfrak{su}(N)$ matrices. Hence, $X_k^\dagger=-X_k$ and $\text{tr}(X_k)=0$. Making use of \myref{surface} and the partition of unity in terms of the projectors ($5$th equation in \myref{ProjecProp}), it can be shown that the algebraic conditions
\begin{equation}
\displaystyle\sum_{k=0}^{N-1}(-1)^kX_k=0 \quad\text{and}\quad \displaystyle\sum_{k=0}^{N-1}X_k=2i\displaystyle\sum_{k=0}^{N-1}\Big(k-\frac{N-1}{2}\Big)P_k,
\end{equation}
hold. This means that the $\mathfrak{su}(N)$-valued immersion functions $X_k$ are linearly dependent. When one multiplies \myref{surface} and \myref{PkXk} by $X_k$ from the left and from the right, respectively, and compares the $P_k\cdot X_k$ obtained from them, one obtains the following cubic matrix equations corresponding to the mixed solutions of \myref{conservation} \cite{Goldstein_Grundland_2}
\begin{equation}\label{equation31}
\left[X_k-ic_k\mathbb{I}_N\right]\left[X_k-i(c_k-1)\mathbb{I}_N\right]\left[X_k-i(c_k-2)\mathbb{I}_N\right]=\varnothing, \qquad 1\leq k\leq N-2.
\end{equation}
In contrast, for the holomorphic $(k=0)$ and the anti-holomorphic $(k=N-1)$ solutions of \myref{conservation}, the constraint relations turn out to be
\begin{alignat}{1}
&\left[X_0-ic_0\mathbb{I}_N\right]\left[X_0-i(c_0-1)\mathbb{I}_N\right] = \varnothing, \label{equation32}\\
&\left[X_{N-1}+ic_{N-1}\mathbb{I}_{N}\right]\left[X_{N-1}+i(c_{N-1}-1)\mathbb{I}_{N}\right] = \varnothing, \label{equation33}
\end{alignat} 
respectively. Since, $X_k$ are anti-Hermitian by construction, one can always diagonalize them, and the Eqs. \myref{equation31}--\myref{equation33}  suggest that the eigenvalues of $X_k$ are
\begin{equation}
ic_k, \quad i\left(c_k-1\right) \quad \text{and}\quad i\left(c_k-2\right), \qquad 0<k<N-1.
\end{equation}
The first two of them stand for the holomorphic solution, while for the anti-holomorphic solution one needs to choose the last two.
\paragraph{Property 2:}
For arbitrary immersion functions $X_k$ of a surface in $\mathfrak{su}(N)$, which are parametrized  in conformal coordinates, satisfy the same EL equation as for the projectors $P_k$ for the $\mathbb{C}P^{N-1}$ model \myref{EulerLagrnage}. That is we have
\begin{equation}
[\partial\bar{\partial}X_k,X_k]=\varnothing.
\end{equation}
\textbf{Proof:}~~~The property was proved in \cite{Goldstein_Grundland_Post}, however, in a slightly complicated way resulting from a different context. Here we provide a simple proof. From \myref{EulerLagrnage} and \myref{TanXP} it is straightforward to obtain $\partial\bar{\partial}X_k=i[\bar{\partial}P_k,\partial P_k]$. Therefore, by expressing $X_k$ in terms of the projectors $P_k$ by using \myref{surface}, we arrive at
\begin{alignat}{1}
[\partial\bar{\partial}X_k,X_k] &= [\bar{\partial}P_k,\partial P_k]\Big(P_k+2\displaystyle\sum_{j=0}^{k-1}P_j\Big)-\Big(P_k+2\displaystyle\sum_{j=0}^{k-1}P_j\Big)[\bar{\partial}P_k,\partial P_k] \\
&= 2[\bar{\partial}P_k,\partial P_k]\displaystyle\sum_{j=0}^{k-1}P_j-2\displaystyle\sum_{j=0}^{k-1}P_j\cdot[\bar{\partial}P_k,\partial P_k],
\end{alignat}
which vanishes according to \myref{equation221} and \myref{equation221a}. \hfill $\Box$
\paragraph{Property 3:}
A straightforward calculation from \myref{surface} yields the Killing form of the two surfaces, i.e. the inner product of the surfaces $X_k$ on $\mathfrak{su}(N)$, as given by
\begin{alignat}{1}
(X_k,X_m) &=-\frac{1}{2}\text{tr}(X_k\cdot X_m)=\frac{Nc_k}{2}(2-c_m), \qquad \forall~m>k, \label{Killing1}\\
(X_k,X_k) &=-\frac{1}{2}\text{tr}(X_k^2)=\frac{Nc_k}{2}(2-c_k)-\frac{1}{2}, \label{Killing2}
\end{alignat}
where we have used the properties \myref{ProjecProp} for the purpose of simplification.
\paragraph{Property 4:}
The product of any number of $X_k$ (including $0$) and at least three of their derivatives $\partial X_k$ always vanishes, i.e.
\begin{equation}\label{equation35}
X_k\cdot\partial X_k\cdot X_k\cdot\partial X_k\cdot\cdot\cdot\cdot\cdot\partial X_k\cdot X_k= X_k\cdot\bar{\partial} X_k\cdot X_k\cdot\bar{\partial}X_k\cdot\cdot\cdot\cdot\cdot\bar{\partial}X_k\cdot\cdot X_k=\varnothing,
\end{equation}
with any arbitrary orderings of $X_k$ and $\partial X_k$, provided that the derivatives are of identical type.\vspace{0.3cm} \\
\textbf{Proof:}~~~We first multiply \myref{surface} by $P_k$ from the left and from the right, so that we obtain
\begin{equation}\label{equation36}
P_k\cdot X_k=X_k\cdot P_k=i\left(c_k-1\right)P_k=P_k\cdot X_k\cdot P_k.
\end{equation}
Now, using \myref{TanXP}, we express the tangent vectors $\partial X_k$ in terms of the projectors $P_k$, and we compute
\begin{alignat}{1}
\partial X_k\cdot\partial X_k &= \partial P_k\cdot P_k\cdot\partial P_k \neq \varnothing, \label{equation37}\\
\partial X_k\cdot\partial X_k\cdot\partial X_k &= i\partial P_k\cdot\partial P_k\cdot P_k\cdot\partial P_k-i\partial P_k\cdot\partial P_k\cdot\partial P_k\cdot P_k. \label{equation38}
\end{alignat}
From \myref{eqn229} we notice that both of the terms on the right hand side of \myref{equation38} are zero. If we multiply more number of $\partial X_k$ in the left hand side of \myref{equation38}, we will end up with similar type of terms in the right hand side but multiplied with more number of $P_k$ and $\partial P_k$, which always vanish due to \myref{eqn229}. Multiplying more number of $X_k$ with \myref{equation38}, we encounter two type of situations, where the products contain either $``\cdot\cdot\cdot P_k\cdot X_k\cdot\cdot\cdot"$ or $``\cdot\cdot\cdot X_k\cdot\cdot\cdot"$. Because of \myref{equation36}, the first case turns out to be zero. In the second case, $X_k$ are multiplied by at least three derivatives and, therefore, by \myref{eqn230} they also vanish. A similar analysis holds for the $\bar{\partial}$ derivatives also. Thus, we prove \myref{equation35}. However, \myref{equation37} implies that the number of derivatives in the product has to be at least three. \hfill $\Box$
\paragraph{Property 5:}
For any number of $X_k$ and their derivatives of identical type (including $0$) with arbitrary ordering, the following traces vanish
\begin{equation}\label{eq31}
\text{tr}\left(X_k\cdot\partial X_k\cdot\partial X_k\cdot X_k\cdot\cdot\cdot\cdot\cdot\partial X_k\right)=\text{tr}\left(X_k\cdot\bar{\partial} X_k\cdot\bar{\partial} X_k\cdot X_k\cdot\cdot\cdot\cdot\cdot\bar{\partial} X_k\right)=0,
\end{equation}
\textbf{Proof:}~~~From \myref{Killing2} we obtain
\begin{equation}\label{ConstTr}
\text{tr}\left(X_k^2\right)=1+Nc_k(c_k-2),
\end{equation}
which implies that 
\begin{equation}\label{TrXDX1}
\text{tr}\left(X_k\cdot\partial X_k\right)=\text{tr}\left(X_k\cdot\bar{\partial} X_k\right)=0.
\end{equation}
Multiplication of $X_k$ by \myref{PkXk} yields $X_k^3$, the trace of which is again a constant, since the trace of the lower order of $X_k$ \myref{ConstTr} is constant. Therefore, by utilizing \myref{TrXDX1}, one obtains $\text{tr}\left(X_k^2\cdot\partial X_k\right)=\text{tr}\left(X_k^2\cdot\bar{\partial} X_k\right)=0$. This process can be continued up to any arbitrary powers of $X_k$, i.e.
\begin{equation}\label{TrXDX}
\text{tr}\left(X_k^n\cdot\partial X_k\right)=\text{tr}\left(X_k^n\cdot\bar{\partial} X_k\right)=0, \qquad n\in\mathbb{Z}^+.
\end{equation}
Now, using \myref{TanXP} we express $\partial X_k$ in terms of the projectors $P_k$, and we compute
\begin{alignat}{1}
\text{tr}\left(\partial X_k\cdot\partial X_k\right) &= 2\text{tr}\left(\partial P_k\cdot\partial P_k\cdot P_k\right), \label{eq35} \\
\text{tr}\left(X_k\cdot\partial X_k\cdot\partial X_k\right) &= \text{tr}\left(\partial X_k\cdot\partial X_k\cdot X_k\right)=2i(c_k-1)\text{tr}\left(\partial P_k\cdot\partial P_k\cdot P_k\right). \label{eq36}
\end{alignat}
The terms on the right hand side of \myref{eq35} and \myref{eq36} vanish due to \myref{eqn229}. In reference to \myref{equation35}, any other cases where the products contain more number of $X_k$'s and their derivatives must vanish. Thus, we prove \myref{eq31}.  \hfill $\Box$
\paragraph{Property 6:}
Differentiating \myref{TrXDX1}, it is easy to show that
\begin{alignat}{1}
\text{tr}\left(X_k\cdot\partial^2 X_k\right)=-\text{tr}\left(\partial X_k\cdot\partial X_k\right) &= 0,\quad \text{tr}\left(X_k\cdot\bar{\partial^2} X_k\right)=-\text{tr}\left(\bar{\partial} X_k\cdot\bar{\partial} X_k\right)=0, \label{eqn35}\\
\text{tr}\left(X_k\cdot\partial\bar{\partial} X_k\right) &= \text{tr}\left(X_k\cdot\bar{\partial}\partial X_k\right)=-\text{tr}\left(\partial X_k\cdot\bar{\partial} X_k\right).
\end{alignat}
In \myref{eqn35}, the traces corresponding to $\partial^2$ and $\bar{\partial^2}$ vanish because of \myref{eq31}. Further differentiations of \myref{eqn35} yield
\begin{equation}
\text{tr}\left(\bar{\partial}\partial X_k\cdot \partial X_k\right)=0, \qquad \text{tr}\left(\partial\bar{\partial} X_k\cdot\bar{\partial} X_k\right)=0.
\end{equation}
From \myref{TanXP}, it is easy to prove that $\bar{\partial}\partial X_k=-i\left[\partial P_k,\bar{\partial}P_k\right]=i\left[\bar{\partial}P_k,\partial P_k\right]=\partial\bar{\partial} X_k$, since $\bar{\partial\partial} P_k=\bar{\partial}\partial P_k$.
\paragraph{Property 7:}
The following traces vanish
\begin{equation}\label{equation319}
\text{tr}\left(\partial\bar{\partial}X_k,\partial^2X_k\right)=0, \qquad \text{tr}\left(\partial\bar{\partial}X_k,\bar{\partial^2}X_k\right)=0.
\end{equation}
\textbf{Proof:}~~~From \myref{TanXP} it is straightforward to calculate the following traces
\begin{alignat}{1}
\text{tr}&\left(\partial\bar{\partial}X_k,\partial^2X_k\right)=\text{tr}\left(\bar{[\partial}P_k,\partial P_k]\cdot[\partial^2P_k,P_k]\right)=\text{tr}\left(P_k\cdot\bar{\partial}P_k\cdot\partial P_k\cdot\partial^2P_k\right. \notag\\ 
& \left.-\bar{\partial}P_k\cdot\partial P_k\cdot P_k\cdot\partial^2P_k+\partial P_k\cdot\bar{\partial} P_k\cdot P_k\cdot\partial^2P_k-P_k\cdot\partial P_k\cdot\bar{\partial} P_k\cdot\partial^2P_k\right)=0,
\end{alignat}
since by \myref{eqn216} one obtains $P_k\cdot\bar{\partial}P_k\cdot\partial P_k=\bar{\partial}P_k\cdot\partial P_k\cdot P_k$ and $P_k\cdot\partial P_k\cdot\bar{\partial} P_k=\partial P_k\cdot\bar{\partial} P_k\cdot P_k$. In a similar way, the other traces in \myref{equation319} can also be shown to be vanished. \hfill $\Box$
\paragraph{Property 8:}
The tangent vectors corresponding to the immersed surfaces take the following forms
\begin{equation}\label{eqn36}
\partial X_k=i\left[\partial X_k,X_k\right], \qquad \bar{\partial} X_k=-i\left[\bar{\partial} X_k,X_k\right].
\end{equation}
\textbf{Proof:}~~~We first collect the terms of \myref{eqn36} from the right hand side to the left hand side and, then, we express $\partial X_k$ and $X_k$ in terms of the projectors $P_k$ by utilizing \myref{PkXk} and \myref{tangent}. Subsequently, we use the orthogonality condition of the projectors, $P_lP_k=\delta_{lk}P_k$ and \myref{comm} to arrive at
\begin{equation}\label{eqn37}
\partial X_k-i\left[\partial X_k,X_k\right]=4i\displaystyle\sum_{j,l=0}^{k-1}\left(\left[P_l,\partial P_j\right]+\left[P_k,\partial P_j\right]\right).
\end{equation}
Next, we replace \myref{eqn37} by \myref{comm}, so that we obtain
\begin{alignat}{1}
\partial X_k-i\left[\partial X_k,X_k\right] &= 4i\partial P_k\cdot\displaystyle\sum_{l=0}^{k-1} P_l-4iP_k\cdot\partial P_k \label{eqn38}\\
&= 4i\left(\partial P_k-\partial P_k\cdot P_k-P_k\cdot \partial P_k\right) =\varnothing, \label{eqn39}
\end{alignat}
which proves \myref{eqn36}. In \myref{eqn38} and \myref{eqn39}, we have used \myref{ProjecProp} and \myref{anticom} whenever required. \hfill $\Box$
\paragraph{Property 9:}
By using \myref{eqn36} successively, we prove
\begin{equation}\label{eqn310}
X_k\cdot\partial X_k\cdot\bar{\partial}X_k\cdot\cdot\cdot\cdot\cdot\partial X_k=\partial X_k\cdot\bar{\partial}X_k\cdot\cdot\cdot\cdot\cdot\partial X_k\cdot\left[X_k+i(M-\bar{M})\mathbb{I}_N\right],
\end{equation}
for any number of derivatives, with $M$ and $\bar{M}$ being the total number of derivatives $\partial X_k$ and $\bar{\partial} X_k$, respectively.  
\paragraph{Property 10:}
In property 4, we proved that the traces of arbitrary number of $X_k$ and their identical derivatives $\partial X_k$ or $\bar{\partial} X_k$ are always zero. Here we provide a more general property for the mixed derivatives, i.e.
\begin{equation}\label{eqn323}
\text{tr}\left(X_k\cdot\partial X_k\cdot X_k\cdot\bar{\partial}X_k\cdot\cdot\cdot\cdot\cdot\partial X_k\right)=0, \qquad M\neq \bar{M}.
\end{equation}
\textbf{Proof:}~~~By taking traces on both sides of \myref{eqn310}, we easily obtain
\begin{equation}\label{eqn324}
\text{tr}\left(\partial X_k\cdot\bar{\partial}X_k\cdot\cdot\cdot\cdot\cdot\partial X_k\right)=0,
\end{equation}
which holds for any type of orderings of the derivatives when $M\neq\bar{M}$. When \myref{eqn324} is multiplied by any number of $X_k$ in arbitrary order, we follow the same argument as given in Property 4 and Property 5 to show that the traces of the corresponding products vanish. When there is one derivative multiplied by any number of $X_k$, because of \myref{TrXDX} the traces are zero. For the cases of three or more derivatives $\partial X_k$ and $\bar{\partial} X_k$ (for $M\neq \bar{M}$) when expressed in terms of the projectors, we obtain the products of $P_k$ and only odd number of $\partial P_k$ in arbitrary orderings. Thus, by following \myref{eqn231}, we prove \myref{eqn323}. \hfill $\Box$
\paragraph{Property 11:}
The following traces are zero
\begin{equation}\label{eq325}
\text{tr}\left(X_k^n\cdot\bar{\partial}X_k\cdot\partial\bar{\partial}X_k\right)=0,\qquad \text{tr}\left(X_k^n\cdot\partial X_k\cdot\bar{\partial}\partial X_k\right)=0, \qquad n\in\mathbb{Z}^+.
\end{equation}
\textbf{Proof:}~~~Let us recall \myref{surface} and \myref{TanXP} to express $X_k,\bar{\partial}X_k$ and $\partial\bar{\partial}X_k$ in terms of the projectors, so that we obtain
\begin{alignat}{1}
\text{tr}\left(X_k\cdot\bar{\partial}X_k\cdot\partial\bar{\partial}X_k\right) &= -i~\text{tr}\Big([c_k\mathbb{I}_N-P_k-2\displaystyle\sum_{j=0}^{k-1}P_j]\cdot[\bar{\partial}P_k,P_k]\cdot[\bar{\partial}P_k,\partial P_k]\Big)\notag \\
&= i(c_k-2)~\text{tr}\left(\bar{\partial}P_k\cdot\partial\bar{\partial}P_k\right), \label{eq326}
\end{alignat}
where we have used \myref{ProjecProp}, \myref{PDPP} and \myref{eqn231} for the purpose of simplification. The right hand side of \myref{eq326} vanishes by virtue of \myref{equation224}. Since, $P_k\cdot X_k=i(c_k-1)P_k$ \myref{equation36}, then \myref{surface} and \myref{PkXk} imply that apart from some different constant terms in front, $X_k^n$ acquires the same terms as $X_k$ has in equation \myref{surface}. Therefore, we obtain 
\begin{equation}
\text{tr}\left(X_k^n\cdot\bar{\partial}X_k\cdot\partial\bar{\partial}X_k\right)=\text{Constant}\cdot\text{tr} \left(\bar{\partial}P_k\cdot\partial\bar{\partial}P_k\right)=0,
\end{equation}
A similar reasoning is applied to the other case in \myref{eq325}. \hfill $\Box$
\paragraph{Property 12:}
For identical type of derivatives, the following relations hold
\begin{alignat}{1}
\partial X_k\cdot\partial X_k\cdot X_k^n &= i^n(c_k-2)^n\partial X_k\cdot\partial X_k= i^n(c_k-2)^n\partial P_k\cdot\partial P_k, \label{eqn331}\\
\bar{\partial} X_k\cdot\bar{\partial} X_k\cdot X_k^n &= i^nc_k^n~\bar{\partial} X_k\cdot\bar{\partial} X_k= i^nc_k^n~\bar{\partial} P_k\cdot\bar{\partial} P_k, \qquad n\in\mathbb{Z}^+. \label{eqn332}
\end{alignat} 
By Property 3, \myref{eqn331} and \myref{eqn332} vanish when more number derivatives are multiplied on the left. For mixed derivatives, we obtain 
\begin{alignat}{1}
\left(\partial X_k\cdot\bar{\partial}X_k\right)^m\cdot X_k^n =& (\alpha^n-\beta^n)\left[\left(\partial X_k\cdot\bar{\partial}X_k\right)^m\cdot X_k^2-2\alpha \left(\partial X_k\cdot\bar{\partial}X_k\right)^m\cdot X_k\right] \label{eq333}\\
& +\left[\beta^n+ic_k\gamma\left(\alpha^n-\beta^n\right)\right]\left(\partial X_k\cdot\bar{\partial}X_k\right)^m, \qquad m=1,2,3,\cdot\cdot\cdot\cdot\cdot \notag \\
\left(\bar{\partial} X_k\cdot\partial X_k\right)^m\cdot X_k^n =& (\alpha^n-\gamma^n)\left[\left(\bar{\partial} X_k\cdot\partial X_k\right)^m\cdot X_k^2-2\alpha \left(\bar{\partial} X_k\cdot\partial X_k\right)^m\cdot X_k\right] \label{eq334}\\
& +\left[\gamma^n+ic_k\gamma(\alpha^n-\gamma^n)\right]\left(\bar{\partial} X_k\cdot\partial X_k\right)^m, \qquad n=0,1,2,\cdot\cdot\cdot\cdot\cdot \notag
\end{alignat}
with $\alpha=i(c_k-1), \beta=ic_k$ and $\gamma=i(c_k-2)$.\\
\textbf{Proof:}~~~The proofs of \myref{eqn331} and \myref{eqn332} follow from simple calculations after representing $X_k,\partial X_k$ and $\bar{\partial}X_k$ in terms of $P_k$ with the help of \myref{surface} and \myref{TanXP}. We first compute
\begin{equation}\label{eqn333}
\partial X_k\cdot\partial X_k=\partial P_k\cdot\partial P_k,
\end{equation}
and, then, we multiply the left hand side of \myref{eqn333} by $X_k$ on the right, so that we obtain
\begin{equation}\label{eqn334}
\partial X_k\cdot\partial X_k\cdot X_k=i(c_k-2)\partial P_k\cdot\partial P_k=i(c_k-2)\partial X_k\cdot\partial X_k.
\end{equation} 
Here we have substituted $X_k$ by $P_k$ from \myref{surface} and simplified it by using \myref{equation220} and \myref{equation221}. When we multiply the left hand side of \myref{eqn334} with more number of $X_k$ on the right, we gain a factor $i(c_k-2)$ each time on the right hand side, so that we arrive at \myref{eqn331} and similarly to \myref{eqn332}. Eqs. \myref{eq333} and \myref{eq334} can be proved in a similar way, however, slightly complicated than before. We compute
\begin{equation}
\left(\partial X_k\cdot\bar{\partial} X_k\right)^m = (-1)^m\left(\partial P_k\cdot\bar{\partial} P_k\right)^m, \qquad m=1,2,3,\cdot\cdot\cdot\cdot\cdot
\end{equation} 
and
\begin{alignat}{1}
\left(\partial X_k\cdot\bar{\partial} X_k\right)^m\cdot X_k^2 &= (-1)^m\left[\beta^2\left(\partial P_k\cdot\bar{\partial} P_k\right)^m+(\alpha^2-\beta^2)\left(\partial P_k\cdot\bar{\partial} P_k\right)^m\cdot P_k\right] \notag\\
&= (-1)^{2m}\left[\beta^2\left(\partial X_k\cdot\bar{\partial} X_k\right)^m+(\alpha^2-\beta^2)\left(\partial X_k\cdot\bar{\partial} X_k\right)^m\cdot P_k\right]\label{eqn339},
\end{alignat}
so that when we substitute $P_k$ from \myref{PkXk} in \myref{eqn339}, we obtain
\begin{alignat}{1}
\left(\partial X_k\cdot\bar{\partial} X_k\right)^m\cdot X_k^2 =& (-1)^{2m}\left[(\alpha^2-\beta^2)\left\{\left(\partial X_k\cdot\bar{\partial} X_k\right)^m\cdot X_k^2-2\alpha \left(\partial X_k\cdot\bar{\partial} X_k\right)^m\cdot X_k\right\}\right. \notag\\
& \left. +\left\{\beta^2-ic_k\gamma(\alpha^2-\beta^2)\right\}\left(\partial X_k\cdot\bar{\partial} X_k\right)^m\right].
\end{alignat}
Computing the higher orders in a similar way, we arrive at \myref{eq333}. \hfill $\Box$
\end{section}
%%%%%%%%%%%%%%%%%%%%%%%%%%%%%%%%%%%%%%%%%%%%%%%%%%%%%%%%%%%%%%%%%%%%%%%%%%%%%%%%%%%%%%%%%%%%%%%%%%%%%%%%%%%%%%%%%%%%%
%  Section 4
%%%%%%%%%%%%%%%%%%%%%%%%%%%%%%%%%%%%%%%%%%%%%%%%%%%%%%%%%%%%%%%%%%%%%%%%%%%%%%%%%%%%%%%%%%%%%%%%%%%%%%%%%%%%%%%%%%%%%
\section{Relation of Weierstrass surface with other integrable surfaces}\label{sec4}
The objective of this section is to realize the connection of generalized Weierstrass representation associated with the $\mathbb{C}P^{N-1}$ model with some other soliton surfaces. To understand the origin of such surfaces, we commence with a brief discussion of the linear spectral problem and, later we establish the consistency of $\mathbb{C}P^{N-1}$ models with the linear spectral problem.
\subsection{$\mathbb{C}P^{N-1}$ model and the linear spectral problem}
Let us summarize briefly the results obtained by Fokas et al. \cite{Fokas_Gelfand_Finkel_Liu} together with those obtained in \cite{Grundland_Post} in order to provide some basic notions for further analysis on $\mathbb{C}P^{N-1}$ models. Suppose that the matrix functions $U_1,U_2\in\mathfrak{g}$ defined on the extended $n$-jet space $\mathcal{N}=(J^n,\lambda)$ satisfy the compatibility condition in two independent variables $u_1$ and $u_2$
\begin{equation}\label{Com}
\frac{\partial U_1}{\partial u_2}-\frac{\partial U_2}{\partial u_1}+[U_1,U_2]=0,
\end{equation} 
which is also known as the zero-curvature condition and, let $\phi$ be a solution of a pair of linear equations
\begin{equation}\label{LaxPair}
\frac{\partial\phi}{\partial u_1}=U_1\phi, \quad \frac{\partial\phi}{\partial u_2}=U_2\phi,
\end{equation}
known as the Lax pair in the literature. Then, the integrable Lax equations \myref{LaxPair} imply that there exists an immersion of the integrable surface $X\in\mathfrak{g}$ of the form \cite{Fokas_Gelfand_Finkel_Liu}
\begin{equation}
X=\tau\phi^{-1}\frac{\partial\phi}{\partial\lambda}+\nu\phi^{-1}(\mbox{pr}(\omega_R\phi)+\phi^{-1}S\phi\in\mathfrak{g}, \quad \tau,\nu\in\mathbb{R},
\end{equation}
with $S$ being an arbitrary $\mathfrak{g}$ valued matrix function and $\mbox{pr}(\omega_R)$ is the prolongation of an evolutionary vector field $\omega_R$ given by
\begin{equation}
\mbox{pr}(\omega_R)=\omega_R+D_JR^k\partial_{u^k_J},\qquad \omega_R=R^k[u]\partial_{u^k}.
\end{equation}
Here, $u^k$ are dependent variables, $u^k_J$ are their derivatives, $D_J$ is the total derivative operator
\begin{equation}
 D_\alpha=\partial_\alpha+u^k_{J,\alpha}\frac{\partial}{\partial u^k_J},\qquad \alpha=1,2,
\end{equation}
$J=(j_1,j_2)$ is the symmetric multi-index and $\omega_R$ is assumed to be a generalized symmetry of the partial differential equation. The cases corresponding to $\nu=S=0$ and $\tau=S=0$ have been studied by Sym-Tafel \cite{Sym,Tafel} and Fokas-Gel'fand \cite{Fokas_Gelfand,Fokas_Gelfand_Finkel_Liu}, respectively, and are known as the Sym-Tafel (ST) and the Fokas-Gel'fand (FG) surfaces in the literature. According to \cite{Grundland_Post}, the FG integrated form of the surfaces associated with the conformal symmetries of the $\mathbb{C}P^{N-1}$ model take the form 
\begin{equation}\label{FGProl}
X^{FG}=\nu\phi^{-1}\mbox{pr}(\omega_R)\phi=\nu\big[f(u_1)\phi^{-1}U_1\phi+g(u_2)\phi^{-1}U_2\phi\big].
\end{equation}
We will study both of the surfaces later in this section. However, before that, let us see how the above scheme fits into the case $\mathbb{C}P^{N-1}$ sigma models. By defining the matrices $U_{1k}$ and $U_{2k}$ in the extended jet space
\begin{equation}\label{MatrixFun}
U_{1k}=\frac{2}{1+\lambda}\left[\partial P_k,P_k\right],\quad U_{2k}=\frac{2}{1-\lambda}\left[\bar{\partial} P_k,P_k\right], \quad U_{1k}^\dagger=-U_{2k},\quad \lambda\in i\mathbb{R},
\end{equation}
the EL equation \myref{EulerLagrnage} become equivalent to \myref{Com} as follows
\begin{equation}\label{EulerLagrnage1}
\left[\partial\bar{\partial}P_k,P_k\right]=\bar{\partial}U_{1k}-\partial U_{2k}+\left[U_{1k},U_{2k}\right]=\varnothing.
\end{equation}
It is, then, possible to write the compatibility conditions in the form of the linear spectral problem
\begin{equation}
\partial\Phi_k=U_{1k}\Phi_k, \qquad \bar{\partial}\Phi_k=U_{2k}\Phi_k,
\end{equation}
so that the wave functions $\Phi_k$ can be integrated for an arbitrary solution of the EL equation \myref{EulerLagrnage1} with finite action as given by \cite{Zakrzewski_Book,Goldstein_Grundland}
\begin{equation}\label{Phi}
\Phi_k=\mathbb{I}_N+\frac{4\lambda}{(1-\lambda)^2}\displaystyle\sum_{j=0}^{k-1}P_j-\frac{2}{1-\lambda}P_k.
\end{equation}
Therefore, the $\mathbb{C}P^{N-1}$ sigma models, indeed, admit a linear spectral problem \cite{Zakrzewski_Book,Zakharov_Mikhailov}. Since, the spectral parameter $\lambda$ is purely imaginary and $\Phi_k$ are the elements of the group $SU(N)$, the inverse of the wave functions can easily be computed as 
\begin{equation}\label{InvPhi}
\Phi_k^{-1}=\mathbb{I}_N-\frac{4\lambda}{(1+\lambda)^2}\displaystyle\sum_{j=0}^{k-1}P_j-\frac{2}{1+\lambda}P_k.
\end{equation}
\subsection{Weierstrass versus Sym-Tafel formulas for immersions}
The ST formula for the immersion of surfaces \cite{Sym,Tafel} as discussed above is given by
\begin{equation}\label{SymOriginal}
X_k^{ST}=\tau\phi_k^{-1}\frac{\partial}{\partial\lambda}\phi_k, \qquad \tau\in\mathbb{R}^+,
\end{equation}
and by definition $X_k^{ST}$ belongs to $\mathfrak{su}(N)$ provided that the matrix functions $U,V$ also belong to $\mathfrak{su}(N)$. However, notice that in the case of $\mathbb{C}P^{N-1}$ models, the matrix functions $U_{1k},U_{2k}$ as defined in \myref{MatrixFun} do not belong to $\mathfrak{su}(N)$. Therefore, for the purpose of making the surface belonging to $\mathfrak{su}(N)$, we adjust the formula of the ST surface \myref{SymOriginal} in the following way
\begin{equation}\label{SymModified}
X_k^{ST}=-i\tau\left(\Phi_k^{-1}\frac{\partial}{\partial\lambda}\Phi_k-\frac{2c_k}{1-\lambda^2}\mathbb{I}_N\right) ~\in\mathfrak{su}(N).
\end{equation}
The purely imaginary factor at the beginning of \myref{SymModified} appears due to the requirement that the matrices $X_k^{ST}$ have to be anti-Hermitian, whereas the additive factor is to ensure that the matrices become traceless. The minus sign at the front is also important for our case, which we will discuss in the following. Making use of \myref{Phi} and \myref{InvPhi}, we obtain the ST formula for the immersion of surfaces \myref{SymModified} in terms of projectors $P_k$ as given by
\begin{equation}\label{Sym}
X_k^{ST}=-\frac{2i\tau}{1-\lambda^2}\Big(P_k+2\displaystyle\sum_{j=0}^{k-1}P_j-c_k\mathbb{I}_N\Big)~\in\mathfrak{su}(N),
\end{equation}  
which coincides with the generalized Weierstrass surface \myref{surface} for $\lambda=\pm\sqrt{1-2\tau}$, with the restrictions that $\lambda$ has singularities at $\pm{1}$. Since, by definition $\tau$ is a positive real number, $\lambda$ always becomes a purely imaginary quantity, which is consistent with the original definition of the spectral parameter coming from the linear spectral problem \myref{MatrixFun}. A positive sign in front of \myref{SymModified} would make $\lambda$ to be real and, therefore, it would destroy the consistency. Nevertheless, what we obtain is a one parameter family of ST surfaces which are related to the Weierstrass surfaces corresponding to the values of the spectral parameter $\lambda$. Of course, when the two surfaces, i.e. the ST and the Weierstrass, merge together, both of the surfaces will automatically satisfy all the properties that we have studied in the previous section.

The algebraic constraints imposed on the surfaces $X_k$ are such that the minimal polynomial \myref{equation31}-\myref{equation33} is of the degree up to three. Let us now find the conditions under which the ST formula for the immersion of surfaces in $\mathfrak{su}(N)$ satisfies the minimal polynomial. For this purpose we substitute the ST formula \myref{Sym} into the relations \myref{equation31}-\myref{equation33}. We start our analysis with the holomorphic solution of the $\mathbb{C}P^{N-1}$ model ($k=0$), with the wave functions given by
\begin{equation}
\Phi_0=\mathbb{I}_N-\frac{2}{1-\lambda}P_0, \qquad \Phi_0^{-1}=\mathbb{I}_N-\frac{2}{1+\lambda}P_0.
\end{equation}
Consequently, the ST surface turns out to be
\begin{equation}\label{X0}
X_0^{ST}=\frac{2i\tau}{1-\lambda^2}\left(c_0\mathbb{I}_N-P_0\right).
\end{equation}
By replacing \myref{X0} into \myref{equation32}, we obtain
\begin{equation}
P_0=\frac{(1-2c_0\tau N-\lambda^2)\left[1-\lambda^2-N(1+2c_0\tau-\lambda^2)\right]}{2\tau N(2-N)(\lambda^2-1)+4\tau^2N^2(2c_0-1)}\mathbb{I}_N.
\end{equation}
Now, we utilize the idempotent property $P_0^2=P_0$ \myref{ProjecProp} to arrive at
\begin{equation}
\frac{(1-2c_0\tau N-\lambda^2)\left[1-\lambda^2-N(1+2c_0\tau-\lambda^2)\right]}{2\tau N(2-N)(\lambda^2-1)+4\tau^2N^2(2c_0-1)}=1,
\end{equation}
which when solved we obtain an algebraic restriction on the spectral parameter
\begin{equation}
\lambda=\pm\sqrt{1-2\tau N(c_0-1)},\quad \pm\sqrt{\frac{N-1+2\tau N(c_0-1)}{N-1}}.
\end{equation}
For the anti-holomorphic solution ($k=N-1$), the wave functions and the ST surface are given by
\begin{equation}
\Phi_{N-1}=\left(\frac{1+\lambda}{1-\lambda}\right)^2\left(\mathbb{I}_{N}-\frac{2}{1+\lambda}P_{N-1}\right), \quad X_{N-1}^{ST}=\frac{2i\tau}{1-\lambda^2}\left[(2+c_{N-1})\mathbb{I}_{N}-P_{N-1}\right],
\end{equation}
respectively. We follow a similar procedure as for the the holomorphic solution to obtain
\begin{equation}
\frac{(1-\lambda^2)^2(1-N)+2\tau N^2\left[(2-c_{N-1})(1-\lambda^2)+2\tau(c_{N-1}+2)^2\right]}{2\tau N^2\left[1+\tau(6+4c_{N-1})-\lambda^2\right]}=1.
\end{equation}
The solutions of which are given by
\begin{equation}
\lambda=\pm\frac{\left[N-1+\tau N^2(c_{N-1}-1)\pm\sqrt{\tau^2N^2\left\{4(N-1)(c_{N-1}+1)^2+N^2(c_{N-1}-1)^2\right\}}\right]^{1/2}}{\sqrt{N-1}}.
\end{equation}
More challenging is the mixed case ($1\leq k\leq N-2$). In this case, we replace $X_k^{ST}$ given by \myref{Sym} into \myref{equation31} and, we obtain the constraint relation as a polynomial in $\lambda$
\begin{equation}
ic_k(c_k-1)(c_k-2)\lambda^6+a_4\lambda^4+a_2\lambda^2+a_0=0,
\end{equation}
with
\begin{eqnarray}
a_4 &=& c_k\left[8-6i+(6+9i)c_k-(2+i)3c_k^2\right]-4, \\
a_2 &=& 8-12i+c_k(2+i)\left[2+20i-(9+6i)c_k+(3-6i)c_k^2\right], \\
a_0 &=& 4+12i-(16+38i)c_k+(38+15i)c_k^2+(2+11i)c_k^3.
\end{eqnarray}
\subsection{Connection between the Fokas-Gel'fand and Weierstrass formulas defined on Minkowski space}
The standard form of the metric in Minkowski coordinates $t,x$ in $\mathbb{R}^2$ is given by
\begin{equation}
dS^2=dt^2-dx^2.
\end{equation}
We shall mainly work on the light-cone coordinates $x^+=t+x,~x^-=t-x$ in $\mathbb{R}^2$ with the metric
\begin{equation}
dS^2=dx^+dx^-.
\end{equation}
We denote the derivatives with respect to $x^+$ and $x^-$ by $\partial_+$ and $\partial_-$, respectively. In what follows, we will refer \cite{Grundland_Snobl} and follow a similar procedure to that discussed in Sec. \ref{sec2} to describe the $\mathbb{C}P^{N-1}$ models in terms of the orthogonal rank-$1$ Hermitian projectors $\mathcal{P}_k$, albeit, functions of the light-cone coordinates $x^+,x^-$. As obviously, the projectors $\mathcal{P}_k$ satisfy all the properties \myref{ProjecProp} as well as those which were explored in Sec. \ref{subsec2.2} with $\partial$ and $\bar{\partial}$ replaced by $\partial_+$ and $\partial_-$, respectively. However, since the partial derivatives $\partial_+\mathcal{P}_k$ and $\partial_-\mathcal{P}_k$ are not complex conjugate to each other, unlike the case of Euclidean space, the Weierstrass formula for the immersion of surfaces in Minkowski space acquires the form
\begin{equation}\label{MinSurface}
\mathcal{X}_k(x^+,x^-)=\int_{\gamma_k(x^+,x^-)}\left(\left[\partial_+\mathcal{P}_k,\mathcal{P}_k\right]dx^+-\left[\partial_- \mathcal{P}_k,\mathcal{P}_k\right]dx^-\right),
\end{equation}
without the multiplicative factor $-i$ when compared to \myref{EuSurface}, which ensures the fact that the immersion of the surface $\mathcal{X}_k(x^+,x^-)$ belongs to $\mathfrak{su}(N)$. In order to write the EL equation \myref{EulerLagrnage1} in a more evolutionary form, we write the matrix functions \myref{MatrixFun} in terms of a set of anti-Hermitian functions $\theta_k=i(\mathcal{P}_k-\mathbb{I}_N/N)\in\mathfrak{su}(N)$ in the following form
\begin{equation}\label{MatrixTheta}
\mathcal{U}_{1k}=-\frac{2}{1+\lambda}[\partial_+\theta_k,\theta_k], \quad \mathcal{U}_{2k}=-\frac{2}{1-\lambda}[\partial_-\theta_k,\theta_k],\quad \mathcal{U}_{1k},\mathcal{U}_{2k}\in\mathfrak{su}(N),
\end{equation}
so that the EL equation \myref{EulerLagrnage1} takes the form
\begin{equation}\label{EulerLagrange2}
[\partial_-\partial_+\theta_k,\theta_k]=\partial_+\mathcal{U}_{2k}-\partial_-\mathcal{U}_{1k}-[\mathcal{U}_{1k},\mathcal{U}_{2k}]=\varnothing.
\end{equation}
Note that, by switching to the Minkowski space the spectral parameter $\lambda$ turns out to be real and, the matrix functions $\mathcal{U}_{1k},\mathcal{U}_{2k}$ defined in \myref{MatrixTheta} become elements of $\mathfrak{su}(N)$. Therefore, the FG formula \myref{FGProl} in the case of $\mathbb{C}P^{N-1}$ models acquires the form
\begin{equation}\label{FGSurface}
\mathcal{X}_k^{FG}=f(x^+)\varPhi^{-1}_k\mathcal{U}_{1k}\varPhi_k+g(x^-)\varPhi^{-1}_k\mathcal{U}_{2k}\varPhi_k \in\mathfrak{su}(N),
\end{equation}
with $\nu=1$ and $\varPhi$ being the solution of the linear spectral problem corresponding to \myref{MatrixTheta}. Let us further consider that $\theta_k$ are the traveling wave solution of the EL equation \myref{EulerLagrange2}, i.e. we express $\theta_k$ as 
\begin{equation}
\theta_k\equiv\theta_k(x^++\kappa x^-), \quad \partial_-\theta_k=\kappa\partial_+\theta_k, \quad \kappa\in\mathbb{R},
\end{equation}
so that the Weierstrass surfaces and their tangent vectors turn out to be
\begin{alignat}{1}
\mathcal{X}_k(x^+,x^-) &=\int_{\gamma_k(x^+,x^-)}\left[\partial_+\theta_k,\theta_k\right](\kappa dx^--dx^+), \\ \partial_+\mathcal{X}_k &=-\left[\partial_+\theta_k,\theta_k\right],\qquad \partial_-\mathcal{X}_k =\kappa\left[\partial_+\theta_k,\theta_k\right]. \label{MinWei}
\end{alignat}
Next, we compute some equivalent relations of \myref{ProjecProp} and \myref{anticom} as
\begin{alignat}{1}
&\mathcal{P}_k^2=\mathcal{P}_k \Rightarrow~~~\theta_k^2=\frac{1-N}{N^2}\mathbb{I}_N+i\frac{N-2}{N}\theta_k, \label{Thetak2}\\
&\{\partial\mathcal{P}_k,\mathcal{P}_k\}=\partial\mathcal{P}_k \Rightarrow~~~\{\partial_+\theta_k,\theta_k\}=i\frac{N-2}{N}\partial_+\theta_k, \label{DTheta}
\end{alignat}
which yield another relation being equivalent to \myref{PDPP} as
\begin{equation}\label{ThetaDTheta}
\mathcal{P}_k\cdot\partial\mathcal{P}_k\cdot\mathcal{P}_k=0 \Rightarrow~~~\theta_k\cdot\partial_+\theta_k\cdot\theta_k=\frac{N-1}{N^2}\partial_+\theta_k.
\end{equation}
Successive uses of \myref{Thetak2}-\myref{ThetaDTheta} produce some important relations
\begin{alignat}{1}
&[\partial_\pm\theta_k,\theta_k]\Big(1+2i\theta_k-\frac{2}{N}\Big)=i\partial_\pm\theta_k, \label{Imp1}\\
&i\Big(1+2i\theta_k-\frac{2}{N}\Big)\partial_\pm\theta_k=[\partial_\pm\theta_k,\theta_k], \label{Imp2} \\
&\big\{[\partial_\pm\theta_k,\theta_k],\partial_\pm\theta_k\big\}=0, \label{Imp3}
\end{alignat}
which we will utilize later. Going back to our discussion, using \myref{MatrixTheta} one can solve the compatibility conditions of the linear spectral problem
\begin{alignat}{1}
\partial_+\varphi_k &=\mathcal{U}_{1k}\varphi_k=-\frac{2}{1+\lambda}[\partial_+\theta_k,\theta_k]\varphi_k, \\ \partial_-\varphi_k &=\mathcal{U}_{2k}\varphi_k=-\frac{2\kappa}{1-\lambda}[\partial_+\theta_k,\theta_k]\varphi_k,
\end{alignat}
to obtain the wave functions in the traveling wave case in the following form \cite{Grundland_Post}
\begin{equation}\label{TravellingWave}
\varphi_k=\Big(\mathbb{I}_N+2i\theta_k-\frac{2}{N}\mathbb{I}_N\Big)e^{2\chi[\partial_+\theta_k,\theta_k]}, \quad \chi=\lambda \Big(\frac{x^+}{1+\lambda}-\frac{\kappa x^-}{1-\lambda}\Big).
\end{equation}
Correspondingly, the form of the FG surface \myref{FGSurface} and its tangent vectors are modified as follows \cite{Grundland_Post}
\begin{alignat}{1}
\mathcal{X}_k^{FG} &=2(\partial_+f-f-\kappa g)\varphi_k^\dagger[\partial_+\theta_k,\theta_k]\varphi_k, \label{FGSurface1}\\
\partial_+\mathcal{X}_k^{FG} &= \Big(\partial_+^2f\chi-\frac{2\partial_+f}{1+\lambda}\Big)\varphi_k^\dagger [\partial_+\theta_k,\theta_k]\varphi_k, \label{TravellingTan} \\
\partial_-\mathcal{X}_k^{FG} &= -\Big(2\partial_-g+\partial_+f\frac{2\kappa\lambda}{1-\lambda}\Big)\varphi_k^\dagger [\partial_+\theta_k,\theta_k]\varphi_k. \label{TravellingTan1}
\end{alignat}
It was argued in \cite{Grundland_Post} that for a general conformal symmetry of the EL equation \myref{EulerLagrnage1}, the immersion function \myref{FGSurface1} in Minkowski space has tangent vectors given by \myref{TravellingTan} and \myref{TravellingTan1} for $f=c_1x^++c_2,g=c_1x^-+c_3$, but not for arbitrary real functions $f$ and $g$. Therefore, in general the surface and its tangent vectors acquire the forms
\begin{alignat}{1}
\mathcal{X}_k^{FG} &=-2(c_1x^++\kappa c_1x^--c_1+c_2+\kappa c_3)\varphi_k^\dagger[\partial_+\theta_k,\theta_k]\varphi_k, \label{FGSurface2}\\
\partial_+\mathcal{X}_k^{FG} &= -\frac{2c_1}{1+\lambda}\varphi_k^\dagger [\partial_+\theta_k,\theta_k]\varphi_k, \quad \partial_-\mathcal{X}_k^{FG} = -2c_1\Big(1+\frac{\kappa\lambda}{1-\lambda}\Big)\varphi_k^\dagger [\partial_+\theta_k,\theta_k]\varphi_k \label{TravellingTan2},
\end{alignat}
with $c_1,c_2$ and $c_3$ being constants. From \myref{TravellingWave}, we now calculate
\begin{alignat}{1}
\varphi_k^\dagger [\partial_+\theta_k,\theta_k]\varphi_k &=\Big(1+2i\theta_k-\frac{2}{N}\Big)e^{-2\chi[\partial_+\theta_k,\theta_k]}[\partial_+\theta_k,\theta_k]\Big(1+2i\theta_k-\frac{2}{N}\Big)e^{2\chi[\partial_+\theta_k,\theta_k]} \notag \\
&=\Big(1+2i\theta_k-\frac{2}{N}\Big)e^{-2\chi[\partial_+\theta_k,\theta_k]}(i\partial_+\theta_k)e^{2\chi[\partial_+\theta_k,\theta_k]} \notag \\
&=i\Big(1+2i\theta_k-\frac{2}{N}\Big)\displaystyle\sum_{n=0}^{\infty}\frac{(-2\chi[\partial_+\theta_k,\theta_k])^n}{n!}(\partial_+\theta_k)e^{2\chi[\partial_+\theta_k,\theta_k]} \notag \\
&=i\Big(1+2i\theta_k-\frac{2}{N}\Big)\partial_+\theta_ke^{4\chi[\partial_+\theta_k,\theta_k]} \notag \\
&=[\partial_+\theta_k,\theta_k]e^{4\chi[\partial_+\theta_k,\theta_k]}, \label{Imp4}
\end{alignat}
where we have used \myref{Imp1}-\myref{Imp3} for the purpose of simplification. We now replace \myref{Imp4} in \myref{TravellingTan2} to obtain
\begin{alignat}{1}
\partial_+\mathcal{X}_k^{FG} &=-\frac{2c_1}{1+\lambda}[\partial_+\theta_k,\theta_k]e^{4\chi[\partial_+\theta_k,\theta_k]},\\
\partial_-\mathcal{X}_k^{FG} &=-2c_1\Big(1+\frac{\kappa\lambda}{1-\lambda}\Big)[\partial_+\theta_k,\theta_k]e^{4\chi[\partial_+\theta_k,\theta_k]},
\end{alignat}
which when compared with \myref{MinWei}, we find a simple constraint relation
\begin{equation}
\kappa=\frac{\lambda^2-1}{\lambda^2+1},
\end{equation}
for which the FG surface coincides with the generalized Weierstrass surface. 
%%%%%%%%%%%%%%%%%%%%%%%%%%%%%%%%%%%%%%%%%%%%%%%%%%%%%%%%%%%%%%%%%%%%%%%%%%%%%%%%%%%%%%%%%%%%%%%%%%%%%%%%%%%%%%%%%%%%%
%  Section 5
%%%%%%%%%%%%%%%%%%%%%%%%%%%%%%%%%%%%%%%%%%%%%%%%%%%%%%%%%%%%%%%%%%%%%%%%%%%%%%%%%%%%%%%%%%%%%%%%%%%%%%%%%%%%%%%%%%%%%
\begin{section}{Conclusions}\label{sec5} 
The main purpose of this paper was to provide a description of $\mathbb{C}P^{N-1}$ sigma models by expressing them in terms of the soliton surfaces immersed in $\mathfrak{su}(N)$ algebra without references to any additional considerations. We provide an extension of the classical Enneper-Weierstrass representation of surfaces in $\mathfrak{su}(N)$ which is identified with the $N^2-1$-dimensional Euclidean space. Several interesting and nontrivial properties of the surfaces in such space have been explored. We have also analyzed the circumstances under which the generalized Weierstrass formula for immersion of $2D$-surfaces coincide with those obtained from the Sym-Tafel formula and the Fokas-Gel'fand formula for immersion. This result allows us to show the equivalence between these three analytic descriptions and to conclude that under certain conditions they parametrized to the same surface in $\mathfrak{su}(N)$.

The proposed approach opens many new directions in which our investigations can be followed up. Firstly, in this paper we have explored some of the properties of the soliton surfaces arising from the $\mathbb{C}P^{N-1}$ sigma models, however, we believe that there are many other interesting properties that remain unexplored. Secondly, it would be exciting to study some other soliton surfaces associated with our system and unify all of them in a single framework. Thirdly, a similar kind of analysis can be carried out for other type of sigma models based on higher rank projectors defined on Grassmannian manifolds or on some homogeneous spaces. Finally and probably the most important problem is to understand the connection of the projector formalism with the discrete integrable models, which are albeit more recent version of the integrable systems. \\

\noindent \textbf{\large{Acknowledgements:}} SD is supported by a CARMIN Postdoctoral Fellowship by IHES and IHP. AMG acknowledges the support of the research grant from NSERC of Canada.

%\begin{subsubsection}*{Acknowledgements:}
%S. D. is supported by the Post Doctoral Fellowship jointly funded by the Laboratory of Mathematical Physics of the %Centre de Recherches Math{\'e}matiques (CRM) and by Prof. Syed Twareque Ali, Prof. Marco Bertola and Prof. %V{\'e}ronique Hussin. The authors thank Prof. Joris Van der Jeugt for his useful discussions.
%\end{subsubsection}     
\end{section}
%%%%%%%%%%%%%%%%%%%%%%%%%%%%%%%%%%%%%%%%%%%%%%%%%%%%%%%%%%%%%%%%%%%%%%%%%%%%%%%%%%%%%%%%%%%%%%%%%%%%%%%%%%%%%%%%%%%%%%

%\bibliographystyle{unsrt}
%\bibliography{CP.bib}

\begin{thebibliography}{100}
\bibitem{Polchinski}
J.~Polchinski,
\newblock {\em String theory}, vol~1,
\newblock Cambridge Univ. Press: New York (1998).

\bibitem{Mccoy_Wu}
B.~M. McCoy and T.~T. Wu,
\newblock {\em The two-dimensional {Ising} model},
\newblock Harvard Uni. Press: Harvard (1973).

\bibitem{Ablowitz}
M.~J. Ablowitz, S.~Chakravarty and R.~G. Halburd,
\newblock Integrable systems and reductions of the self-dual {Yang--Mills} equations,
\newblock {J. Math. Phys.} \textbf{44}, 3147--3173 (2003).

\bibitem{Nelson}
D.~Nelson, T.~Piran and S.~Weinberg,
\newblock {\em Statistical mechanics of membranes and surfaces},
\newblock World Scientific: Singapore (2004).

\bibitem{Rozhdestvenski}
B.~L. Rozhdestvenskii and N.~N. Janenko,
\newblock {\em Systems of quasilinear equations and their applications to gas dynamics}, Math. monograph vol~55,
\newblock AMS: Providence (1983).

\bibitem{David_Ginsparg}
F.~David, P.~Ginsparg and J.~Zinn-Justin,
\newblock {\em Fluctuating geometries in statistical mechanics and field theory},
\newblock Elsevier: Amsterdam (1996).

\bibitem{Lipowsky}
R.~Lipowsky and E.~Sackmann,
\newblock {\em Structure and dynamics of membranes},
\newblock Elsevier: Amsterdam, 1995.

\bibitem{Davydov}
A.~S. Davydov,
\newblock Solitons in molecular systems,
\newblock {Phys. scr.} \textbf{20}, 387 (1979).

\bibitem{Zhong}
O.~Zhong-Can, L.~Ji-Xing and X.~Yu-Zhang,
\newblock {\em Geometric methods in the elastic theory of membranes in liquid crystal phases},
\newblock World Scientific: Singapore (1999).

\bibitem{Landolfi}
G.~Landolfi,
\newblock New results on the {Canham--Helfrich membrane model via the generalized Weierstrass representation},
\newblock {J. Phys. A: Math. Gen.} \textbf{36}, 11937 (2003).

\bibitem{Safran}
S.~A Safran,
\newblock {\em Statistical thermodynamics of surfaces, interfaces and membranes},
\newblock Addison-Wesley: New York (1994).

\bibitem{Goldstein_Grundland}
P.~P. Goldstein and A.~M. Grundland,
\newblock Invariant recurrence relations {for $\mathbb{C}P^{N-1}$ models},
\newblock {J. Phys. A: Math. Theor.} \textbf{43}, 265206 (2010).

\bibitem{Din_Zakrzewski}
A.~M. Din and W.~J. Zakrzewski,
\newblock General classical solutions in {the $\mathbb{C}P^{N-1}$ model},
\newblock {Nuc. Phys. B} \textbf{174}, 397--406 (1980).

\bibitem{Manton_Sutcliffe_Book}
N.~Manton and P.~Sutcliffe,
\newblock {\em Topological solitons},
\newblock Cambridge Univ. Press: New York (2004).

\bibitem{Zakharov_Mikhailov}
V.~E. Zakharov and A.~V. Mikhailov,
\newblock Relativistically invariant two-dimensional models of field theory which are integrable by means of the inverse scattering problem method,
\newblock {Sov. Phys. JETP} \textbf{47}, 1017--1027 (1978).

\bibitem{Grundland_Strasburger_Zakrzewski}
A.~M. Grundland, A.~Strasburger and W.~J. Zakrzewski,
\newblock Surfaces immersed in {$\mathfrak{su}(N+1)$ Lie algebras obtained from the $\mathbb{C}P^{N-1}$ sigma models},
\newblock {J. Phys. A: Math. Gen.} \textbf{39}, 9187 (2006).

\bibitem{Grundland_Yurdusen}
A.~M. Grundland and {\.I}.~Yurdu{\c{s}}en,
\newblock On analytic descriptions of two-dimensional surfaces associated with {the $\mathbb{C}P^{N-1}$ sigma} model,
\newblock {J. Phys. A: Math. Theor.} \textbf{42}, 172001 (2009).

\bibitem{Nomizu_Sasaki_Book}
K.~Nomizu and T.~Sasaki,
\newblock {\em Affine differential geometry: geometry of affine immersions},
\newblock Cambridge Univ. Press: Cambridge, England (1994).

\bibitem{Konopelchenko}
B.~G. Konopelchenko,
\newblock Induced surfaces and their integrable dynamics,
\newblock {Stud. Appl. Math.} \textbf{96}, 9--51 (1996).

\bibitem{Konopelchenko_Taimanov}
B.~G. Konopelchenko and I.~A. Taimanov,
\newblock Constant mean curvature surfaces via an integrable dynamical system,
\newblock {J. Phys. A: Math. Gen.} \textbf{29}, 1261 (1996).

\bibitem{Goldstein_Grundland_1}
P.~P. Goldstein and A.~M. Grundland,
\newblock Invariant description of {$\mathbb{C}P^{N-1}$} sigma models,
\newblock {Theor. Math. Phys.} \textbf{168}, 939--950 (2011).

\bibitem{Goldstein_Grundland_2}
P.~P. Goldstein and A.~M. Grundland,
\newblock On the surfaces associated with {$\mathbb{C}P^{N-1}$} models,
\newblock {JPCS} \textbf{284}, 012031 (2011).

\bibitem{Goldstein_Grundland_Post}
P.~P. Goldstein, A.~M. Grundland and S.~Post,
\newblock Soliton surfaces associated with sigma models: differential and algebraic aspects,
\newblock {J. Phys. A: Math. Theor.} \textbf{45}, 395208 (2012).

\bibitem{Fokas_Gelfand_Finkel_Liu}
A.~S. Fokas, I.~M. Gel'fand, F.~Finkel and Q.~M. Liu,
\newblock A formula for constructing infinitely many surfaces on {Lie} algebras and integrable equations,
\newblock {Sel. Math.} \textbf{6}, 347--375 (2000).

\bibitem{Grundland_Post}
A.~M. Grundland and S.~Post,
\newblock Soliton surfaces associated with generalized symmetries of integrable equations,
\newblock {J. Phys. A: Math. Theor.} \textbf{44}, 165203 (2011).

\bibitem{Sym}
A.~Sym,
\newblock Soliton surfaces and their applications (soliton geometry from spectral problems),
\newblock In {\em Geometric Aspects of the Einstein Equations and Integrable Systems}
\newblock (Lecture Notes in Physics, vol 239), pages 154--231, Springer: Berlin (1985).

\bibitem{Tafel}
J.~Tafel,
\newblock Surfaces in $\mathbb{R}^3$ with prescribed curvature,
\newblock {J. Geome. Phys.} \textbf{17}, 381--390 (1995).

\bibitem{Fokas_Gelfand}
A.~S. Fokas and I.~M. Gelfand,
\newblock Surfaces on {Lie groups, on Lie algebras, and their integrability},
\newblock {Commun. Math. Phys.} \textbf{177}, 203--220 (1996).

\bibitem{Zakrzewski_Book}
W.~J. Zakrzewski,
\newblock {\em Low dimensional sigma models},
\newblock Adam Hilger: Bristol (1989).

\bibitem{Grundland_Snobl}
A.~M. Grundland and L.~{\v{S}}nobl,
\newblock Description of surfaces associated with {Grassmannian sigma models on Minkowski space},
\newblock {J. Math. Phys.} \textbf{46}, 083508 (2005).
\end{thebibliography}

\end{document}